# New Consideration on Composed Nonextensive Magnetic Systems

F. A. R. Navarro[(a)], M. S. Reis[(b)], E. K Lenzi[(c)] and I. S. Olivera[(a)]

(a) Centro Brasilero de Pesquisa Físicas, Rua Dr. Xavier Sigaud 150, Rio de Janeiro, 22290-180, Brazil.

(b) Departamento de Física and CICECO, Universidade de Aveiro 3810-193 Aveiro, Portugal.

(c) Departamento de Física, Universidade Estadual de Maringa, Av. Colombo 5790, 87020-900, Maringa, PR, Brazil.

## Abstract

In this paper a composed $A+B$ magnetic system, with spins $J_A=2$ and $J_B=3/2$, is considered within the mean-field approximation, in the framework of Tsallis nonextensive statistics. Our motivation is twofold: (1) to approach the existing experimental data of manganese oxides (manganites), where $Mn^{3+}$ and $Mn^{4+}$ form two magnetic sublattices, and (2) to investigate the structure of nonextensive density matrices of composed systems. By imposing that thermodynamic quantities, such as the magnetization of sublattices $A$ and $B$, must be invariant weather the calculation is taken over the total Hilbert space or over partial subspaces, we found that the expression for the nonextensive entropy must be adapted. Our argument is supported by calculation of sublattices magnetization $M_A$ and $M_B$, internal energy, $U_A$ and $U_B$, and magnetic specific heat, $C_A$ and $C_B$. It is shown that only with the modified entropy the two methods of calculation agree to each other. Internal energy and magnetization are additive, but no clear relationship was found between $S_A$, $S_B$ and the total entropy $S_{A+B}$ for $q \neq 1$. It is shown that the reason for the failure of the standard way of calculation is the assumption of statistical independence between the two subsystems, which however does not affect the density matrix in the full Hilbert space.

## 1. Introduction

Recently, Reis *et. al.* [1, 2, 3] have provided experimental evidences that the magnetic properties of manganese oxides, or simply manganites, must be interpreted within the framework of Tsallis non-extensive statistics [4, 5]. This proposal should interest both, the magnetic and the statistical physics communities, for at least two reasons: (1) up to date no first-principle model is known to account for the whole complexity of the physical properties of manganites. Therefore, phenomenological tools to treat these materials should be welcome; (2) samples of manganites are simple to be made, and even sophisticated high purity single crystal samples are nowadays available [6]. Therefore, from one hand manganites are an easy way to test the consequences of nonextensive statistics and, from another, nonextensive statistics is a parametrical tool to treat manganites. One of the greatest challenges of the theoretical nonextensive framework is to find relationships between the entropic parameter $q$ and the dynamical variables of a system. In Ref. [3] it is analytically shown that $q$ relates to the magnetic susceptibility of the material. Finally, one must mention that manganites present many of the correct ingredients for nonextensivity: they are fractal [6], they are inhomogeneous [7, 8] and they exhibit long-range interaction [6].

In a previous publication [9] we investigated the magnetic properties of a nonextensive composed system of two spins 1/2 within the mean-field approximation. It was shown that in order the sublattices magnetization, calculated from the full Hilbert space, to agree with the same quantity calculated from the subspaces, the respective partial traces should be performed over the quantity $\rho^q$ instead of $\rho$. This conflicts with the standard approach, and raises important questions concerning the structure of nonextensive density matrix space, particularly the notion of statistical independence. Besides, $\rho^q$ cannot be interpreted as a true density matrix, since it is not normalized. In the present paper, we carried the analysis further and show that with an appropriate modification in the definition of the nonextensive entropy, it is possible to keep the usual interpretation of partial tracing, at the same time making the two methods of calculation agree to each other. Such a modification does not affect the usual formalism in any fashion, whenever full Hilbert space is under consideration. The starting point is the simple remark that thermodynamic observables of a composed system must not depend on the dimension of Hilbert spaces. Therefore, we impose the following guiding principle for a composed $A+B$ system,

$$O_f = O_p, \qquad (1)$$

where $O$ is a thermodynamic quantity of either subsystem, corresponding to a quantum mechanical observable $\hat{O}$. The subscripts $f$ and $p$ stand for *full* and *partial* Hilbert spaces, respectively.

## 2. Model

Similarly to what was done in Ref. [10], we propose that nonextensive entropy must be defined as:

$$S_q = -k \frac{Tr[\rho^q Ln(\rho^q)]}{Tr(\rho^q)}, \qquad (2)$$

and the *q*-logarithm [4, 5]:

$$Ln_q(\rho) = \frac{\rho^{1-q} - I}{1-q}, \qquad (3)$$

where *I* is the identity matrix in the full Hilbert space. From (2) and (3) we obtain the normalized form of the nonextensive entropy [11]:

$$S_q = \frac{1 - Tr(\rho^q)}{(q-1)Tr(\rho^q)}. \qquad (4)$$

Maximization of $S_q$ in Eq. (4), subject to the norm and energy constraints:

$$Tr(\rho) = 1 \quad \text{and} \quad U_q = \frac{Tr(\rho^q H)}{Tr(\rho^q)}, \qquad (5)$$

leads to the well known equilibrium distribution:

$$\rho = \frac{\{1 - (1-q)\beta' H\}^{\frac{1}{1-q}}}{Z_q}, \qquad (6)$$

where $Z_q$ is the generalized partition function:

$$Z_q = Tr\{1 - (1-q)\beta' H\}^{\frac{1}{1-q}} \qquad (7)$$

and

$$\beta' = \frac{\beta}{\frac{1}{Tr(\rho^q)} + (1-q)\beta U_q} \qquad (8)$$

according to this, if $\rho$ describes a composed *A+B* system, each subsystem entropy should be given, in the full Hilbert space, by:

$$S_A = -k \frac{Tr[\rho^q Ln(\rho'_A)]}{Tr(\rho^q)} \quad \text{and} \quad S_B = -k \frac{Tr[\rho^q Ln(\rho'_B)]}{Tr(\rho^q)}, \qquad (9)$$

where:

$$\rho'_A = \rho_A \otimes I_B \quad \text{and} \quad \rho'_B = I_A \otimes \rho_B, \qquad (10)$$

where $I_A$ and $I_B$ are the identity matrices in the respective Hilbert subspaces, and $\rho_A = Tr_B(\rho)$ and $\rho_B = Tr_A(\rho)$.

Upon the *assumption* of statistical independence,

$$\rho_{A+B}^q = \rho_A^q \otimes \rho_B^q, \tag{11}$$

the standard formulae for the entropies in the subspaces are obtained. We denominated this the method I:

$$S_A = \frac{1 - Tr(\rho_A^q)}{(q-1)Tr(\rho_A^q)} \quad \text{and} \quad S_B = \frac{1 - Tr(\rho_B^q)}{(q-1)Tr(\rho_B^q)}, \tag{12}$$

which relates to the total entropy, $S_{A+B}$ through the formula:

$$S_{A+B} = \frac{S_A}{Tr(\rho_B^q)} + \frac{S_B}{Tr(\rho_A^q)} + (1-q)S_A S_B \tag{13}$$

However, in order to preserve the traditional and well established notions of partial traces and expected values, we propose that the correct way to calculate the subsystems entropy (in the subspaces $A$ and $B$) must include weight factors $E_A$ and $E_B$. We call this procedure the method II:

$$S_A = -k\frac{Tr[E_A Ln(\rho_A)]}{Tr(E_A)} \quad \text{and} \quad S_B = -k\frac{Tr[E_B Ln(\rho_B)]}{Tr(E_B)}, \tag{14}$$

where

$$E_A = Tr_B(\rho^q) \quad \text{and} \quad E_B = Tr_A(\rho^q). \tag{15}$$

However, it is simple to see that $F_A$ and $F_B$ cannot be taken as the density matrices of the subsystems, for if they were, one would have:

$$Tr_A(E_A) \equiv Tr_{A+B}(\rho^q) \quad \text{and} \quad Tr_B(E_B) \equiv Tr_{A+B}(\rho^q), \tag{16}$$

from which, using the standard formulae, we would arrive to the following inconsistencies:

$$S_{A+B} \equiv S_A \quad \text{and} \quad S_{A+B} \equiv S_B. \tag{17}$$

The $F$ factors can be also included in the definition of the non-normalized form of Tsallis entropy. This will affect only the expression for the energy parameter $\beta'$ [2].

We will now exemplify the above considerations with the case of a system composed by two magnetically coupled sublattices, $A$ and $B$, with spins $J_A=2$ and $J_B=3/2$. These are the spins of free ion $Mn^{3+}$ and $Mn^{4+}$, respectively. In the mean-field approximation, the Hamiltonian is given by [12]:

$$H_M = H_A + H_B, \tag{18}$$

with

$$H_A = -g_A J_A^z B_A \quad \text{and} \quad H_B = -g_B J_B^z B_B, \quad (19)$$

where the effective field acting one each subsystem is:

$$B_A = B_0 + \lambda_A M_A + \lambda_{AB} M_B \quad \text{and} \quad B_B = B_0 + \lambda_B M_B + \lambda_{AB} M_A, (20)$$

where $B_0$ is the external field and $g_A=2/3$ and $g_B=4/5$. The intralattice couplings satisfy: $\lambda_A, \lambda_B \rangle 0$, and the interlattices coupling satisfy: $\lambda_{AB} \rangle 0$ for ferromagnetic and $\lambda_{AB} \langle 0$ for antiferromagnetic ordering [13].

In the present case, the dimension of the full Hilbert space is *20x20*, and the Hilbert subspaces of the subsystems are *5x5* ($J^A=2$) and *4x4* ($J^B=3/2$) dimensional.

In the usual formalism, method I, we have that the sublattices magnetizations are given by:

$$M_A = g_A \frac{Tr(\rho_A^q J_A^z)}{Tr(\rho_A^q)} \quad \text{and} \quad M_B = g_B \frac{Tr(\rho_B^q J_B^z)}{Tr(\rho_B^q)}, \quad (21)$$

where

$$\rho_{A(B)}^q = \{Tr_{B(A)}(\rho)\}^q ; \quad (22)$$

whereas according to our proposal (method II), they are given by:

$$M_A = g_A \frac{Tr(E_A J_A^z)}{Tr(E_A)} \quad \text{and} \quad M_B = g_B \frac{Tr(E_B J_B^z)}{Tr(E_B)}, \quad (23)$$

In the full Hilbert space we obviously have:

$$M_{A(B)} = g_{A(B)} \frac{Tr(\rho^q J_{A(B)}^z)}{Tr(\rho^q)}. \quad (24)$$

Similarly, the contribution of each sublattice to the internal energy are:

$$U_A = \frac{Tr(E_A H_A)}{Tr(E_A)} \quad \text{and} \quad U_B = \frac{Tr(E_B H_B)}{Tr(E_B)}, \quad (25)$$

and to the specific heat [14]:

$$C_A = T\left\{\frac{\partial S_A}{\partial T}\right\}_{B_0} \quad \text{and} \quad C_B = T\left\{\frac{\partial S_B}{\partial T}\right\}_{B_0}. \quad (26)$$

Notice that, for $q \neq 1$, the nonadditivity of the entropies leads to nonadditivity of the specific heat. The above quantities were calculated self-consistently using the two approaches, the standard one and our proposal.

## 3. Computational Simulations

The Figure 1 shows the comparison between the temperature dependence of the magnetization of sublattices *A* and *B*, calculated in the full Hilbert space with method I, Eq. (21), and method II, Eq. (23). The two methods are equivalents only for values of *q* close to 1. Therefore the Boltzmann-Gibbs statistics do not present troubles at different spaces Hilbert.

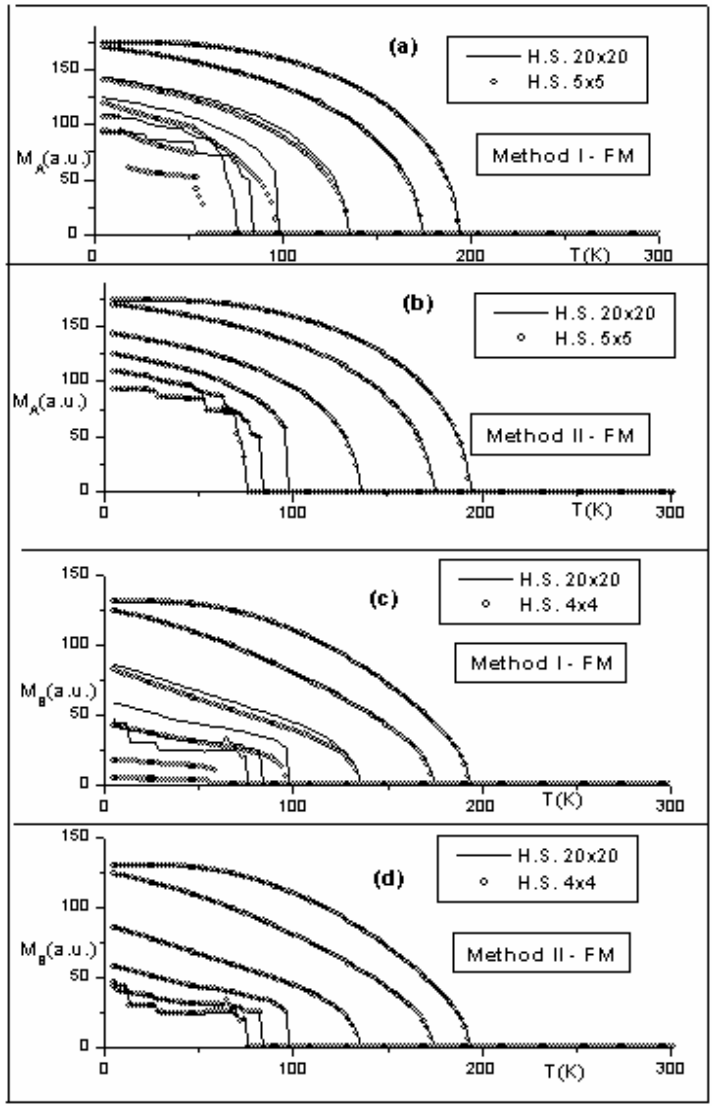

Fig. 1. Magnetizations of sublattices *A* and *B* of a ferromagnetically coupled system. Continuous line is the calculation made in the full Hilbert space, and symbols are the results of calculation in subspaces using either method I or II Notice that only method II reaches plain agreement with the calculation in full Hilbert space. The values of *q*, from down to upper, are 0.1, 0.3, 0.5, 0.7, 0.9 and 1.0. This remark is valid to all the figures.

The Figure 2 shows the similar results for the internal energies $U_A$ and $U_B$. Moreover, only method II reaches plain agreement with the calculation in full Hilbert space. The values of $q$, from up to down, are 0.1, 0.3, 0.5, 0.7, 0.9 and 1.0.

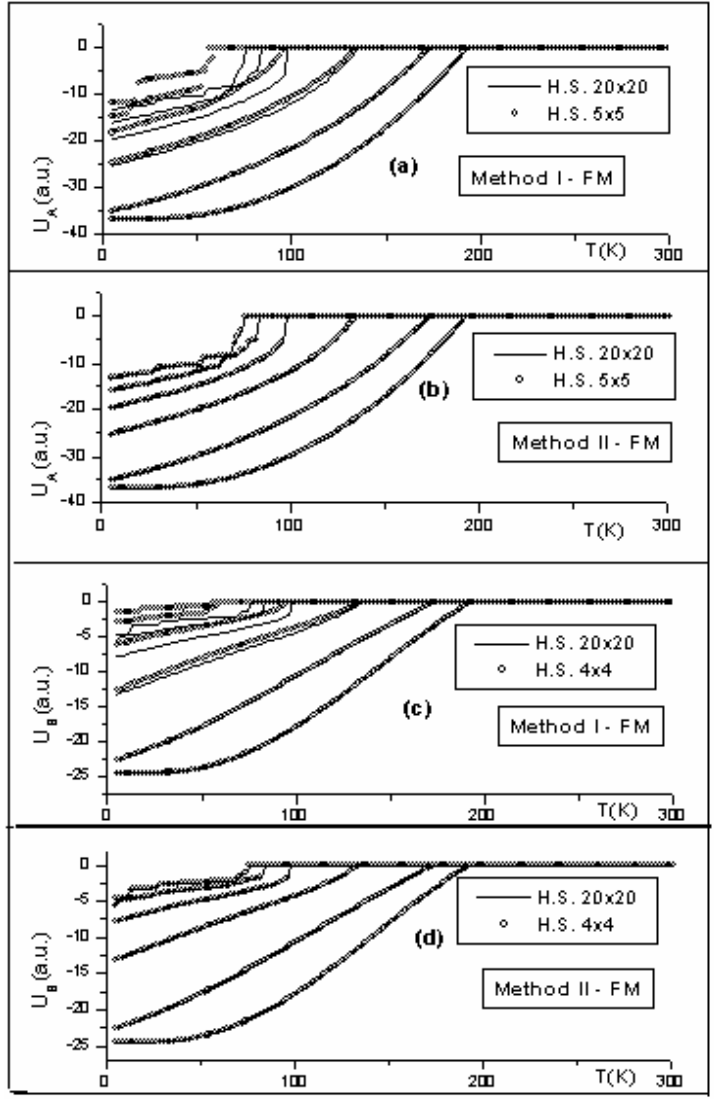

Fig. 2. Internal energies of sublattices $A$ and $B$ of a ferromagnetically coupled system. Continuous line is the calculation made in the full Hilbert space, and symbols are the results of calculation in subspaces using either method I or II

The Figure 3 compares the contributions to the specific heat. The values of $q$, from down to upper, are 0.1, 0.3, 0.5, 0.7, 0.9 and 1.0.

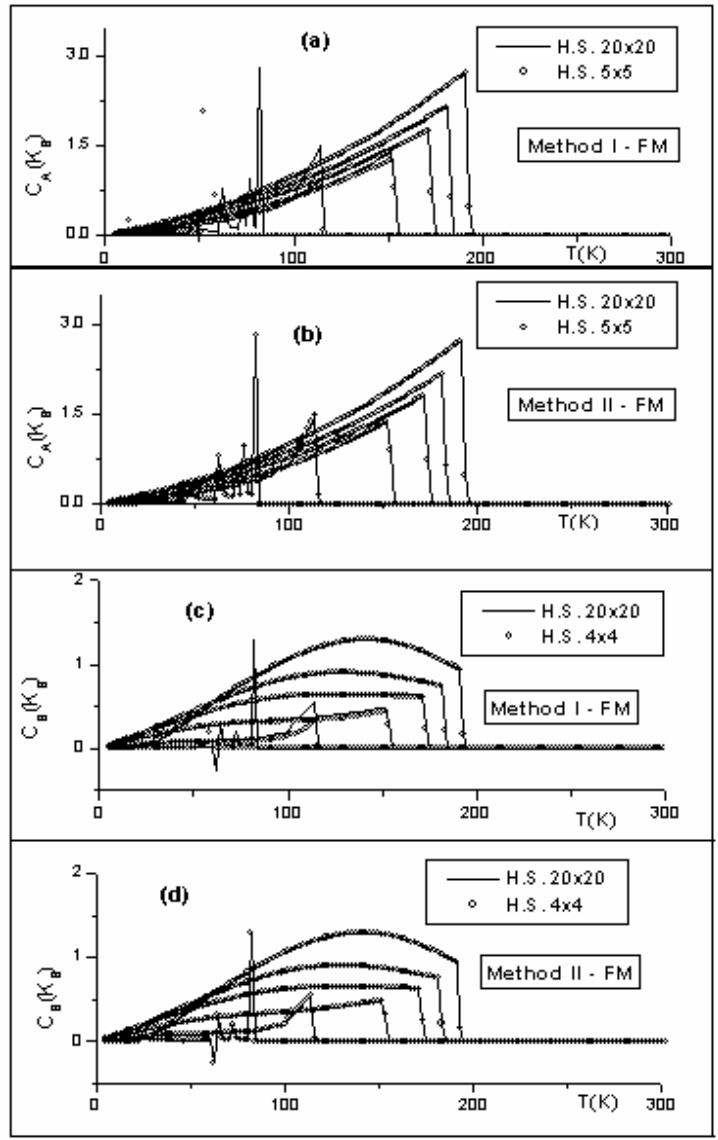

Fig. 3. Contribution to the magnetic specific heat from sublattices *A* and *B* of a ferromagnetically coupled system. Continuous line is the calculation made in the full Hilbert space, and symbols are the results of calculation in subspaces using either method I or II. This latter method II is correct.

The Figure 4 shows that nonextensive magnetizations and internal energy become additive only on method II. Calculation was made for $q$ values 0.3 and 0.7.

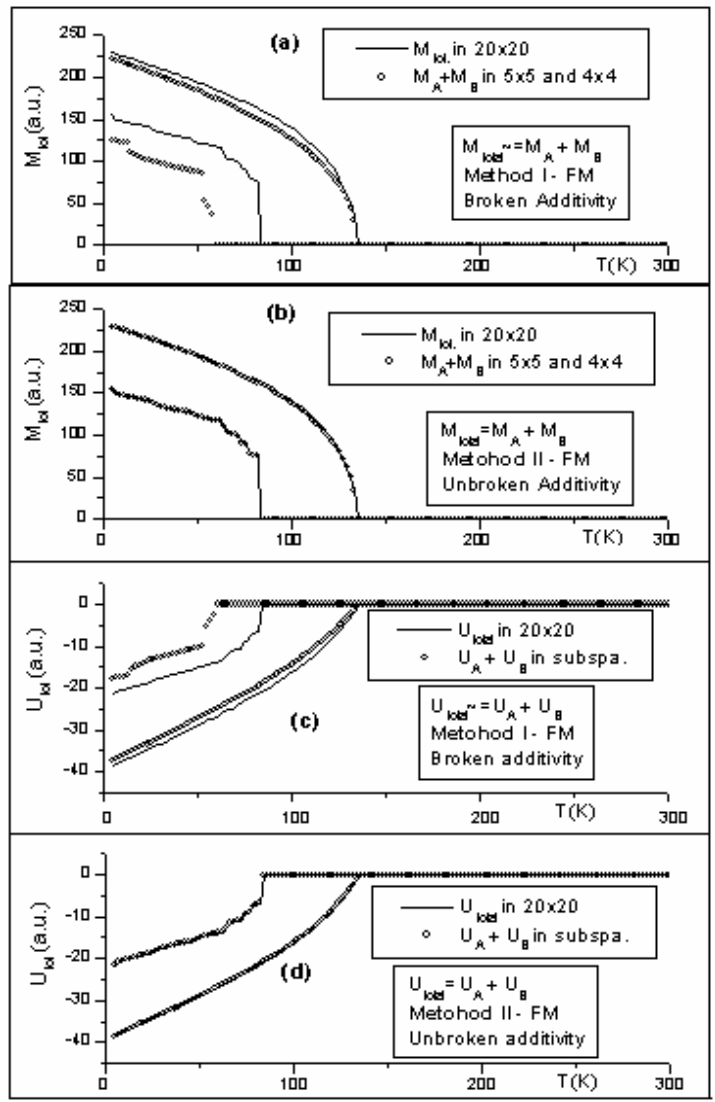

Fig. 4 The additivity in the nonextensive formalism occurs only on method II of calculation. This is a desired property for a magnetic system, since that it preserves the total number of spins.

## 4. Conclusions

In conclusion, one could not derive an analytical expression relating $S_A$ and $S_B$ to $S_{A+B}$. The reason for the failure of the standard method of calculation (method I), Eqs. (21) (and similar expressions for other quantities) is that it is valid only for statistically independent systems, a notion which is still not clear in the context of nonextensive statistics, and of course does not apply to the case of coupled magnetic systems. On the other hand, the standard expressions can be derived from the expressions (23) and (25), which can then be considered a generalization of the first ones. These observations apply to the entropy, as well. Therefore, we conclude that Eqs. (14) must replace the usual expressions, Eqs. (12) for the entropy of nonextensive composed systems. The analysis of the nonextensive entropy of composed systems and applications in systems such as alloys, glasses and magnetoresistive materials [14, 15, 16]. This is an important aspect to the formalism of nonextensive statistics, but which has not been enough developed. From one hand, the possibility of making an experimental connection to manganites opens up the possibility of laboratorial tests of the formalism, and from another hand offers to the magnetism community a convenient parametric tool to interpret and classify the magnetic behaviour of manganites.